\documentclass[groupedaddress, aps, longbibliography, twocolumn, prl]{revtex4-1}

\usepackage{graphicx}
\usepackage{bm}
\usepackage{bbold}
\usepackage{physics}
\usepackage{xcolor}
\usepackage{enumitem}
\usepackage{amsmath, amssymb}
\usepackage[normalem]{ulem}
\usepackage{natbib}
\usepackage{physics}

\newcommand{\rhoin}{\ensuremath{\rho_\text{in}}}
\newcommand{\rhoout}{\ensuremath{\rho_\text{out}}}
\newcommand{\eye}{\ensuremath{\mathbb{1}}}
\newcommand{\swap}{\ensuremath{\textsf{SWAP}}}
\renewcommand{\bell}{\ensuremath{B^+}}

\usepackage[colorlinks=true]{hyperref}
\hypersetup{citecolor = blue}

\begin{document}

\title{Postselection-free entanglement dynamics via spacetime duality}

\author{Matteo Ippoliti}
\affiliation{Department of Physics, Stanford University, Stanford, CA 94305, USA}
\author{Vedika Khemani}
\affiliation{Department of Physics, Stanford University, Stanford, CA 94305, USA}

\begin{abstract}
    The dynamics of entanglement in `hybrid' non-unitary circuits (for example, involving both unitary gates and quantum measurements) has recently become an object of intense study. 
    A major hurdle toward experimentally realizing this physics is the need to apply \emph{postselection} on random measurement outcomes in order to repeatedly prepare a given output state, resulting in an exponential overhead.
    We propose a method to sidestep this issue in a wide class of non-unitary circuits by taking advantage of \emph{spacetime duality}.
    This method maps the purification dynamics of a mixed state under non-unitary evolution onto a particular correlation function in an associated unitary circuit. 
    This translates to an operational protocol which could be straightforwardly implemented on a digital quantum simulator. 
    We discuss the signatures of different entanglement phases, and demonstrate examples via numerical simulations. 
    With minor modifications, the proposed protocol allows measurement of the purity of arbitrary subsystems, which could shed light on the properties of the quantum error correcting code formed by the mixed phase in this class of hybrid dynamics. 
\end{abstract}

\maketitle


The dynamics of quantum entanglement is a topical area of research in several subfields of physics ranging from quantum information and quantum gravity to condensed matter and atomic physics~\cite{Horodecki2009, Fazio2008, Eisert2010, Calabrese2009, Bardarson2012, kimhuse, Hartman2013, tsunami, Mezei2017, Nahum2017, VonKeyserlingk2018, Huang2019, Brown2019}. 
Recent works have begun to extend this line of research to \emph{non-unitary} settings, involving many-body systems subject to repeated measurements~\cite{Li2018, Skinner2019, Li2019, Choi2020, Gullans2020PRX, Fan2020, Gullans2020PRL, Bao2020, Jian2020, Cao2019, NahumSkinner2020, Zabalo2020, Ippoliti2020, Lavasani2020, Hsieh2020, Tang2020, LopezPiqueres2020, Li2020, Turkeshi2020, Alberton2020, Fuji2020, Pal2020, Romito2020, Vijay2020, LiFisher2020, Fidkowski2020, NahumRoy2020}. 
Instead of averaging over measurement outcomes, one can consider individual quantum trajectories corresponding to particular sequences of measurement outcomes~\cite{dalibard1992wave}. Such ``monitored dynamics" are an essential feature of near-term quantum devices in which modulated interactions with an environment, say via measurements, are necessary for unitary control and feedback. Remarkably, one finds that the steady state ensembles of monitored dynamics display novel entanglement phases and phase transitions. 
These include area-law and volume-law entangled phases separated by critical points described by conformal field theories.
Many questions, on both the transitions and the phases themselves, remain active areas of study---notably the universality class of the transitions~\cite{Bao2020, Jian2020, Li2020} and the nature of the volume-law phase, which is understood as a dynamically-generated quantum error-correcting code (QECC) hiding information from local measurements~\cite{Choi2020, Gullans2020PRX, Fan2020, Ippoliti2020, LiFisher2020}.

Measuring entanglement generally requires the preparation of many identical copies of the same state (either simultaneously or sequentially)~\cite{Horodecki2003, Pichler2012, Abanin2012, Beenakker2012, Elben2018, Elben2019, Elben2020, You2020}.
In the presence of measurements, this becomes extremely challenging as it requires \emph{postselection}:
a quantum measurement is an intrinsically random process whose outcomes are sampled stochastically with Born probabilities, and a quantum trajectory in this evolution is associated with a \emph{specific} record of measurement outcomes. 
Hence, preparing multiple copies of the same state incurs an exponential postselection overhead $e^{O(pLT)}$ in the size $L$ and depth $T$ of the circuit (assuming a finite rate of measurement $p$). 
There are ways to partly overcome this challenge: 
(i) using a single reference qubit as a \emph{local probe} of the entanglement phase~\cite{Gullans2020PRL}, which considerably alleviates the postselection overhead; 
(ii) in Clifford circuits, using a combination of classical simulations and feedback to force specific measurement outcomes by error-correcting any ``wrong'' ones.
Nevertheless, it is still desirable to \emph{directly} access the entanglement properties of more general non-unitary and non-Clifford evolutions. 

In this Letter, we propose a novel method to access quantum entanglement in a broad class of non-unitary circuits \emph{without} facing an exponential postselection barrier. 
Specifically, we will consider non-unitary circuits that are \emph{`spacetime dual'} (explained below) to unitary evolutions; and we propose a method for measuring the purity $\Tr (\rho^2)$, related to the second Renyi entropy via $\Tr(\rho^2) = e^{-S_2(\rho)}$, for the whole system as well as for arbitrary subsystems.
This allows access to the purification dynamics of an initially mixed state, which is intimately related to the entanglement dynamics of pure states~\cite{Gullans2020PRX}. 
In particular, the volume-law entangled phase maps onto the \emph{mixed} phase, in which the dynamics generates a QECC that protects information from measurements, so that an initially mixed state remains mixed for exponentially long times.
Subsystem purity measurements contain key information about the nature of this as-of-yet poorly understood QECC~\cite{LiFisher2020, Gullans2020PRX, Ippoliti2020}.

We note that while the class of non-unitary circuits for which our method applies is not completely general, it still encompasses a vast space---as large as the space of unitary circuits built from local two-qubit gates. In particular it applies to circuits with (a specific class of) unitary gates interspersed with (specific types of) forced projective measurements.


\begin{figure}
    \centering
    \includegraphics[width=\columnwidth]{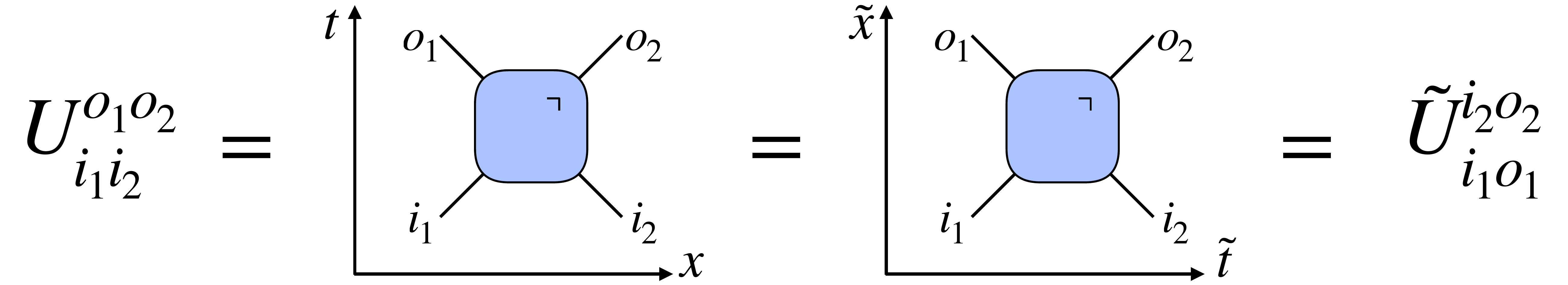}
    \caption{Spacetime duality. By swapping the spatial and temporal axes, a unitary gate $U$ maps onto another, generally non-unitary gate $\tilde{U}$.}
    \label{fig:flipgate}
\end{figure}

{\it Spacetime duality.}
Given a two-qubit unitary gate $U_{i_1, i_2}^{o_1, o_2}$, mapping input qubits $i_{1,2}$ (bottom legs) to output qubits $o_{1,2}$ (top legs), we define its spacetime dual $\tilde{U}$ as the matrix obtained by flipping the arrow of time by 90 degrees, i.e. viewing the left legs as inputs and the right legs as outputs: $U_{i_1, i_2}^{o_1, o_2} \equiv \tilde{U}_{i_1,o_1}^{i_2, o_2}$ (Fig.~\ref{fig:flipgate}).
The result of this transformation, $\tilde{U}$, is generally not unitary (gates $U$ such that $\tilde{U}$ is also unitary are known as ``dual-unitary" and have been studied intensely recently~\cite{Bertini2018PRL, Bertini2019PRX, Gopalakrishnan2020, Bertini2019PRL, Piroli2020, Claeys2020, Kos2020, Klobas2020}).
The generic non-unitarity of $\tilde{U}$, and the possibility that it might counter entanglement growth, has also been employed to pursue more efficient tensor network contraction schemes~\cite{Cirac2009, Hastings2014}, and similar ideas we applied to study the complexity of shallow (2+1)-dimensional circuits~\cite{Harrow2020}.

In general, one has $\tilde{U} \equiv 2VH$, where $V$ is unitary and $H$ is a positive semidefinite matrix of unit norm, which can be seen as a generalized measurement (i.e. an element of a POVM set~\cite{NielsenChuangBook}, see~\footnote{See online supplemental material for additional details on the spacetime duality transformation, the whole-system and subsystem purity measurement protocols, the state preparation protocol, and the quantum code properties in the mixed phase.} for more details).
As an example, $U = \eye$ yields $\tilde{U} = 2\ket{\bell} \bra{\bell}$, where $\ket{\bell} = (\ket{00} + \ket{11})/\sqrt{2}$ is a Bell pair state. 
Thus the spacetime duality transformation generally maps unitary circuits to non-unitary \emph{hybrid circuits} involving unitary gates as well as (weak or projective) measurements, up to prefactors. 
Crucially, the measurements are \emph{forced}: the outcome is deterministic; no quantum randomness is involved. In the example of $U=\eye$, the outcome is always $\ket{\bell}$. The ability to avoid postselection in our protocol stems from this observation.


\begin{figure}
    \centering
    \includegraphics[width=\columnwidth]{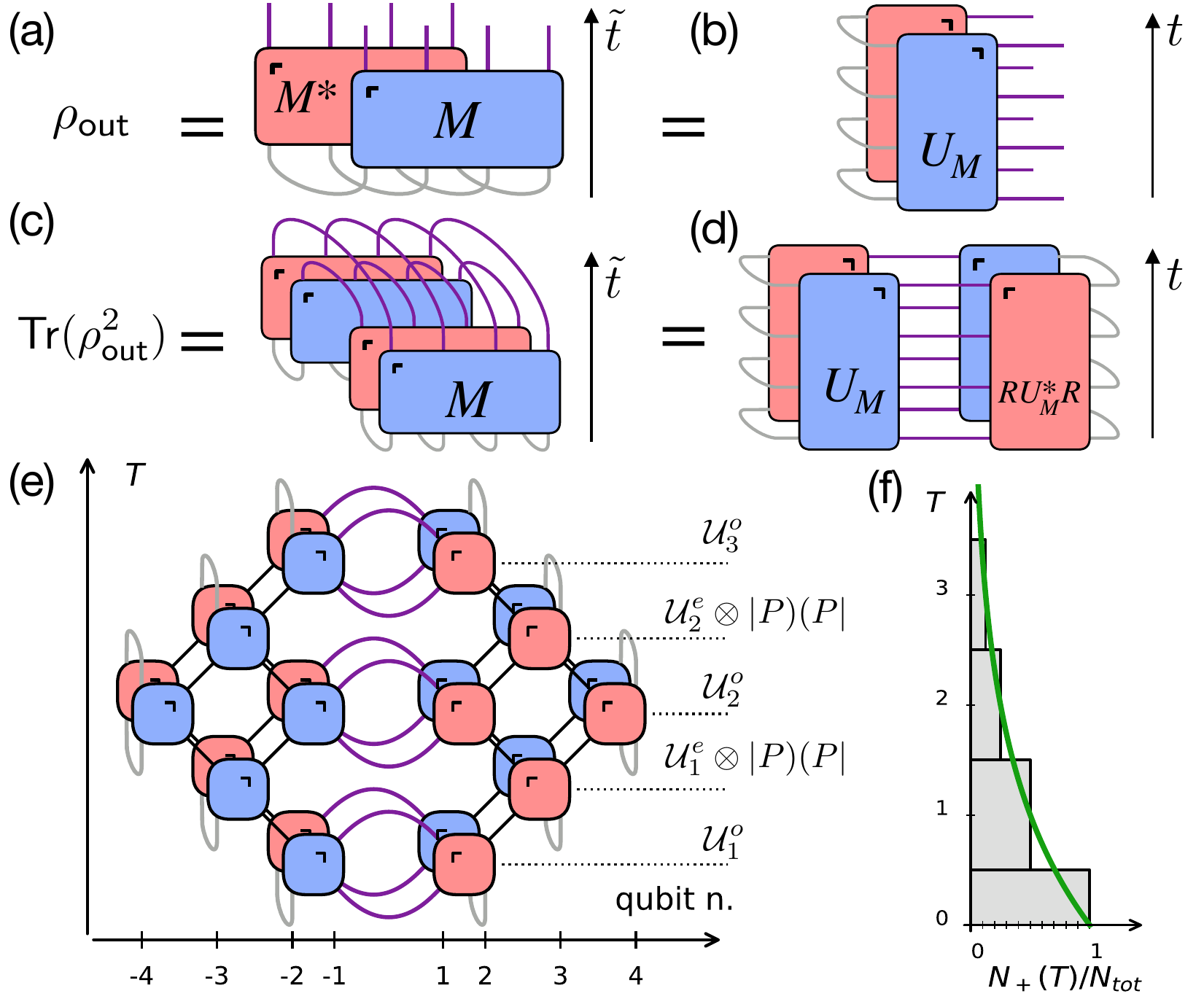}    \caption{(a) Purification dynamics: a fully mixed state $\rho_{\text{in}} \propto\eye$ (gray lines, bottom) is evolved by a hybrid circuit $M$, yielding a state $\rho_{\text{out}} \propto MM^\dagger$ (purple lines, top).
    (b) If $M = \tilde{U}_M$ with $U_M$ unitary, the purification process is `spacetime-dual' to a unitary evolution. 
    $\rhoout$ lives on a time-like slice of the circuit (purple legs, right).
    (c) Purity of the output state, $\text{Tr} (\rho_\text{out}^2)$.
    (d) Same tensor network viewed as a correlation function in a unitary circuit.
    The arrow of time $\tilde{t}$ ($t$) denotes hybrid (unitary) evolution.
    (e) Protocol for measuring the purity of $\rhoout$, sketched for $T=3$. 
    Tensor network legs are color-coded as in (a-d). Qubits $\pm 1$ are repeatedly initialized in the Bell state state $P=\ket{\bell}\bra{\bell}$ (upward purple arcs) and measured in the Bell basis (downward purple arcs); the protocol succeeds if all $T$ Bell measurements yield $\ket{\bell}$.
    (f)
    The fraction of runs that are successful up to time $T$, $N_+(T)/N_\text{tot}$, yields the purity of $\rho_\text{out}$ on $\tilde{L}=2T$ qubits.} 
    \label{fig:diagram}
\end{figure}

{\it Postselection-free measurement protocol.}
The idea is to use a ``laboratory'' system, whose evolution is unitary, to simulate a ``dual'' system whose evolution is non-unitary and realizes purification dynamics. We will use $t$ ($\tilde{t}$) do denote the arrow of time in the unitary (non-unitary) evolution.
The target purification dynamics starts with a fully mixed state, $\rhoin = \eye/2^{\tilde{L}}$ on an even number $\tilde{L}$ of qubits (we use a tilde for quantities defined in the non-unitary time direction). This is evolved by a non-unitary circuit including forced measurements, $M$. 
The output state, $\rhoout \propto M M^\dagger$ (Fig.~\ref{fig:diagram}(a)), may be partially or completely purified. 
We focus on non-unitary circuits $M$ whose spacetime dual is a unitary circuit, i.e. $M = \tilde{U}_M$ with $U_M$ unitary; in this case, it is possible to view $\rhoout$ as living on a time-like slice at the spatial edge of a unitary circuit (Fig.~\ref{fig:diagram}(b)). 
Likewise the purity, $\Tr(\rhoout^2)$ (Fig.~\ref{fig:diagram}(c)), maps under spacetime duality to a (multi-point) correlation function in an associated unitary evolution (Fig.~\ref{fig:diagram}(d)): a bipartite one-dimensional chain in which the left half evolves under the unitary $U_M$, the right half evolves under $RU_M^\ast R$ ($R$ denotes spatial inversion), and the only region where the evolution is non-unitary is the central pair of qubits, where horizontal bonds (space-like qubit worldlines) implement the product of $\rhoout$ with itself. 
We additionally note that unitarity of $U_M$ (coupled with special {\it `depolarizing'} boundary conditions, discussed in detail in~\cite{Note1}) elides all gates outside forward and backward lightcones emanating from the central pair of qubits, as in Fig.~\ref{fig:diagram}(e). 

The goal of the following discussion is to provide an interpretation of the non-unitary processes taking place at the central bond, so that this tensor contraction can be converted into an operational prescription for the measurement of the purity. 
In the ``laboratory'' (unitary) time direction, one has a chain of $L \equiv \tilde{L}+2$ qubits evolved for time $T\equiv \tilde{L}/2$.
We symmetrically label the qubits as $i = \{\pm 1, \pm 2, \dots \pm (T+1)\}$, and denote qubits $i \leq -2$ as $\mathcal L$ (left), $i=\pm 1$ as $\mathcal C$ (central), and $i\geq 2$ as $\mathcal R$ (right). 
The system is initialized in the state $\rho \propto \eye_{\mathcal L} \otimes P_{\mathcal C} \otimes \eye_{\mathcal R}$, where $P = \ket{\bell}\bra{\bell}$ and $\ket{\bell} = \frac{1}{\sqrt{2}}(\ket{00}+\ket{11})$ is a Bell state. It is then evolved in time by a circuit with a brickwork structure. 
First a layer of unitary gates, represented by super-operator $\mathcal{U}_t^o$, acts on the `odd' bonds---a layer of the circuit $U_M$ on $\mathcal{L}\cup\{-1\}$ and a layer of $RU_M^\ast R$ on $\mathcal{R}\cup\{+1\}$. 
Next, a similarly-defined unitary layer $\mathcal{U}_t^e$ acts on `even' bonds, which do not include $\mathcal{C}$. 
There, a \emph{forced} Bell measurement takes place: 
$\rho\mapsto P_{\mathcal C} \otimes \Tr_{\mathcal C}(P_{\mathcal C} \rho)$. 
The process terminates at time $t=T$ with a final forced measurement of $P_{\mathcal C}$.  Using the operator-state representation, in which $|A)$ denotes an operator $A$ as a state with inner product $(A|B) = \Tr A^\dagger B$, the forced Bell measurement reads $|\rho) \mapsto |P_{\mathcal C})(P_{\mathcal C}|\rho)$, and the overall evolution can be written as
\begin{align}
    \Tr(\rhoout^2)
    & \propto (P| \mathcal{U}^o_T \circ \prod_{t=1}^{T-1}  \left[ \mathcal{U}_t^e \otimes |P_{\mathcal C})(P_{\mathcal C}| \right] \circ \mathcal{U}_t^o |P) \;,
    \label{eq:purity_to_correlation}
\end{align}
where $|P)\equiv |P_{\mathcal C}) \otimes |\eye_{\mathcal{L} \cup \mathcal{R}})$.
The purity of the hybrid circuit output $\rhoout$ is thus mapped to a $(2T)$-point correlation function of the projector $P_{\mathcal C}$ during a unitary evolution obtained from the original hybrid circuit $M$ via the spacetime duality.

We are now in a position to recast the result in Eq.~\eqref{eq:purity_to_correlation} as an operational protocol for measuring the purity of the state of interest, $\rhoout$.
The protocol consists of the following steps:
{\bf (1)} Choose an integer $T$ and prepare a $2(T+1)$-qubit chain in the state $\rho = \eye_{\mathcal L} \otimes P_{\mathcal C} \otimes \eye_{\mathcal R} / 4^T$. Set $t=1$.
{\bf (2)} Evolve odd bonds unitarily under $\mathcal U_t^o$.
{\bf (3)} Perform a Bell measurement on $\mathcal C$. If the outcome is $\ket{\bell}$, continue; otherwise, stop and record a {\bf failure}.
{\bf (4)} If $t=T$, stop and record a {\bf success}; otherwise, evolve even bonds unitarily under $\mathcal{U}_t^e$, increment $t$ by one, and go back to step 2.
Let the number of ``successful'' runs out of $N_\text{tot}$ trials be $N_+(T)$; then, 
\begin{equation}
    \Tr({\rhoout^2}|_{\tilde{L} = 2T, \tilde{T}=T}) = N_+(T)/ N_\text{tot} \;,
    \label{eq:protocol}
\end{equation}
where $\rhoout|_{\tilde{L}=2T, \tilde{T}=T}$ denotes the output state of hybrid dynamics on a system of $\tilde{L} = 2T$ qubits evolved for time $\tilde{T} = T$. 
This is the central result of this Letter.
Note that while we have not kept track of numerical prefactors in this derivation, the proportionality constant in Eq.~\eqref{eq:protocol} turns out to be exactly one, see~\cite{Note1}.

We emphasize that the above protocol, despite featuring projective measurements and runs ending in ``failure'', does \emph{not} use postselection.
Indeed, ``failures'' increment the denominator $N_\text{tot}$ in Eq.~\eqref{eq:protocol} and provide crucial information for the purity measurement.
In other words, copies of the same state $\rhoout$ (identical up to control errors) can be realized deterministically arbitrarily many times. 
The exponential overhead of postselection is entirely removed. 
Finally we remark that if one wants to prepare $\rhoout$ in real space (as opposed to a time-like slice of the circuit as obtained above), this can be achieved with an approach based on gate teleportation~\cite{NielsenChuangBook}, using $2T$ ancillas initialized in $\ket{\bell}$ Bell states and $O(T^2)$ $\swap$ gates in a 1D geometry, see~\cite{Note1}.


{\it Entanglement phases.}
While a typical hybrid circuit is generally \emph{not} dual to a unitary circuit, the space of models we address is still very large, and it is reasonable to expect a rich variety of purification phases and entanglement phenomena within this class of models.  Here we begin to explore this space for the purpose of demonstrating that interesting purification phases are indeed possible, while leaving the longer-term enterprise of charting this space to future work.

Surprisingly, despite the presence of measurements, Eq.~\eqref{eq:protocol} suggests that the generic outcome of the purification dynamics in these models should be a mixed phase, in which $\rhoout$ has extensive entropy. Indeed, if the late-time probability of obtaining $\ket{\bell}$ as the outcome of the Bell measurement on $\mathcal C$ approaches any value $p_\infty < 1$, then $N_+(T)$---which requires the outcome of \emph{all} $T$ Bell measurements to be $\ket{\bell}$---decays exponentially at late times. Therefore the state has a finite entropy density $s_2$ directly measurable from a decay time constant:
\begin{equation}
\Tr(\rhoout^2|_{\tilde{L} = 2T, \tilde{T}=T}) \sim e^{-T/\tau}
\implies 
s_2 \equiv (2\tau)^{-1} \;.
\label{eq:entropy_density}
\end{equation}
This is another main result of this Letter.

The mixed-phase outcome should be expected whenever the unitary circuit $U_M$ features any amount of scrambling: then, any projector $\ket{\bell} \bra{\bell}$ injected in $\mathcal C$ will irreversibly grow into a global operator, never refocusing at $\mathcal C$; thus the probability of later obtaining $\ket{\bell}$ as a Bell measurement outcome will be lower than 1, and the above argument will give a mixed phase.
However, exceptions are possible in non-scrambling circuits.

\begin{figure}
    \centering
    \includegraphics[width=\columnwidth]{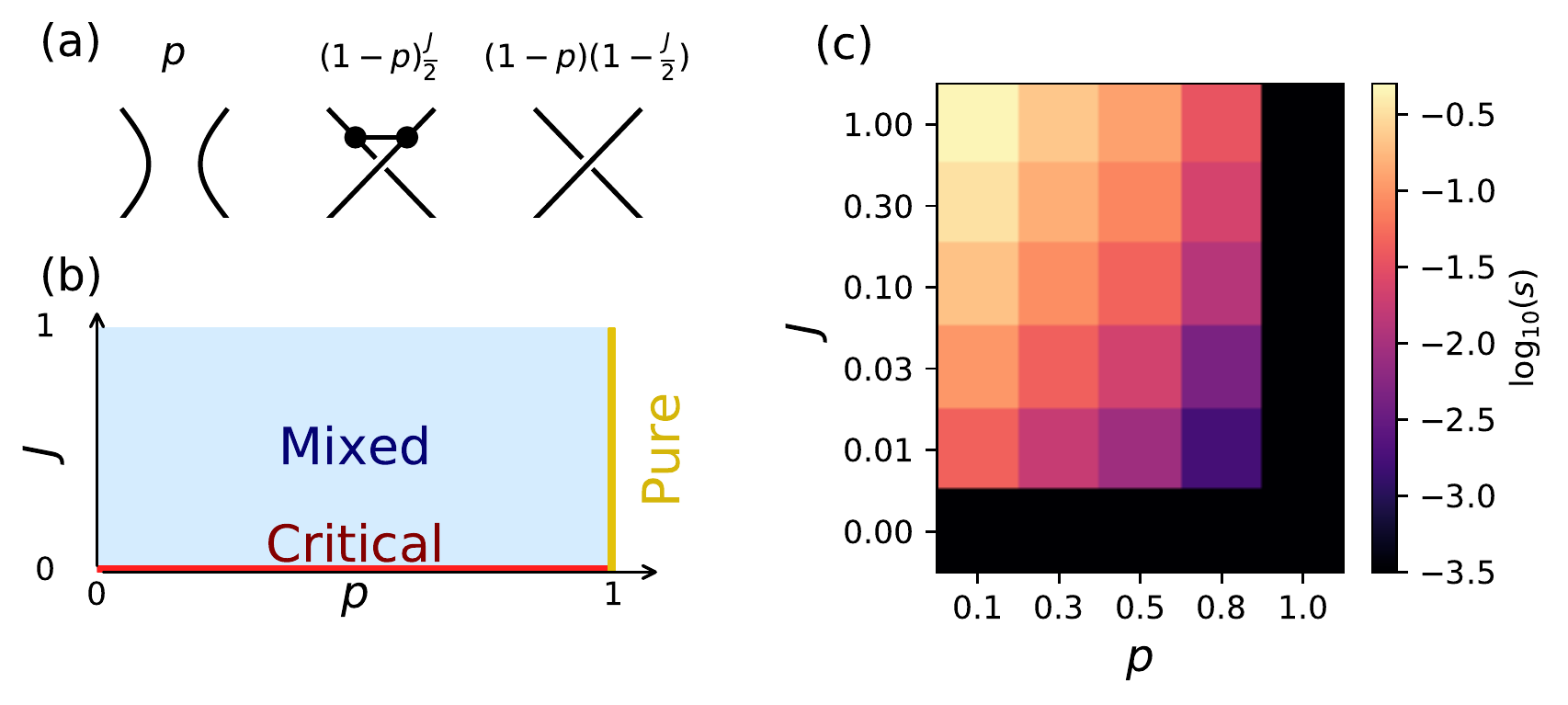}
    \caption{(a) Summary of the Clifford circuit model: probabilities of $\eye$, $\textsf{iSWAP}$ and $\swap$ gates. 
    (b) Schematic of purification phase diagram of the dualized circuit vs $p$, $J$. 
    (c) Entropy density of hybrid circuit output state $\rhoout$ vs $p$, $J$ (Clifford simulations on $\tilde L \leq 4096$ qubits).
    }
    \label{fig:phase_diagram}
\end{figure}

As an illustration, we map out the purification phases for a specific model.  For computational simplicity and closer comparison with the known phenomenology, we choose a set of unitary Clifford circuits whose spacetime duals consist only of unitary gates and projective Pauli measurements.
We consider a brickwork layer of two-qubit Clifford gates chosen as indicated in Fig.~\ref{fig:phase_diagram}(a): $\text{Prob}(\eye) = p$, $\text{Prob}(\textsf{iSWAP}) = (1-p)J/2$~\footnote{The \textsf{iSWAP} gate is defined as $e^{-i \frac{\pi}{4}(XX+YY)}$, and is equal to the product of a $\swap$ and a Clifford $ZZ$ interaction $e^{i\frac{\pi}{4}ZZ}$.}, $\text{Prob}(\swap)=(1-p)(1-J/2)$. Arbitrary single-qubit Clifford gates act before and after each two-qubit gate so there are no symmetries in the model.
As $\swap$ and $\textsf{iSWAP}$ are dual-unitary while $\tilde{\eye} \propto \ket{\bell} \bra{\bell}$ is a projector, $p\in [0,1]$ serves as the measurement rate for the dual hybrid circuit.
$J\in[0,1]$ serves as an interaction rate, with $J=0$ giving a non-interacting `swap circuit'.
We note that the spacetime dual of this unitary model is not too different from the original unitary-projective model considered in Ref.~\cite{Li2018}: single-site $Z$ measurements are replaced by two-qubit Bell measurements; gates are sampled out of exactly half of the two-qubit Clifford group, rather than the whole group. 
Remarkably, these seemingly small changes yield a completely different phase diagram, sketched in Fig.~\ref{fig:phase_diagram}(b).

Via stabilizer numerical simulations we find three possible outcomes across the $(p,J)$ parameter space (results for the entropy density are shown in Fig.~\ref{fig:phase_diagram}(c)).
A pure phase is only possible at $p=1$, where the circuit $U_M$ consists purely of identity gates and $\rhoout = (\ket{\bell}\bra{\bell})^{\otimes T}$ is trivially a pure state. 
Remarkably, it is unstable to infinitesimal perturbations away from $p=1$, in sharp contrast to other unitary-projective models where the pure phase is generic for sufficiently high measurement rates.
For any $J>0$ and $p<1$ (i.e. almost all of parameter space) we have a mixed phase: indeed, this is the default outcome for generic interacting circuits.
Finally, on the $J=0$ (and $0<p<1$) line, we have a critical purification phase, with vanishing entropy density $s_2=0$ but divergent entropy $S_2\sim \sqrt{T}$, which we characterize in the following.

\begin{figure}
    \centering
    \includegraphics[width=\columnwidth]{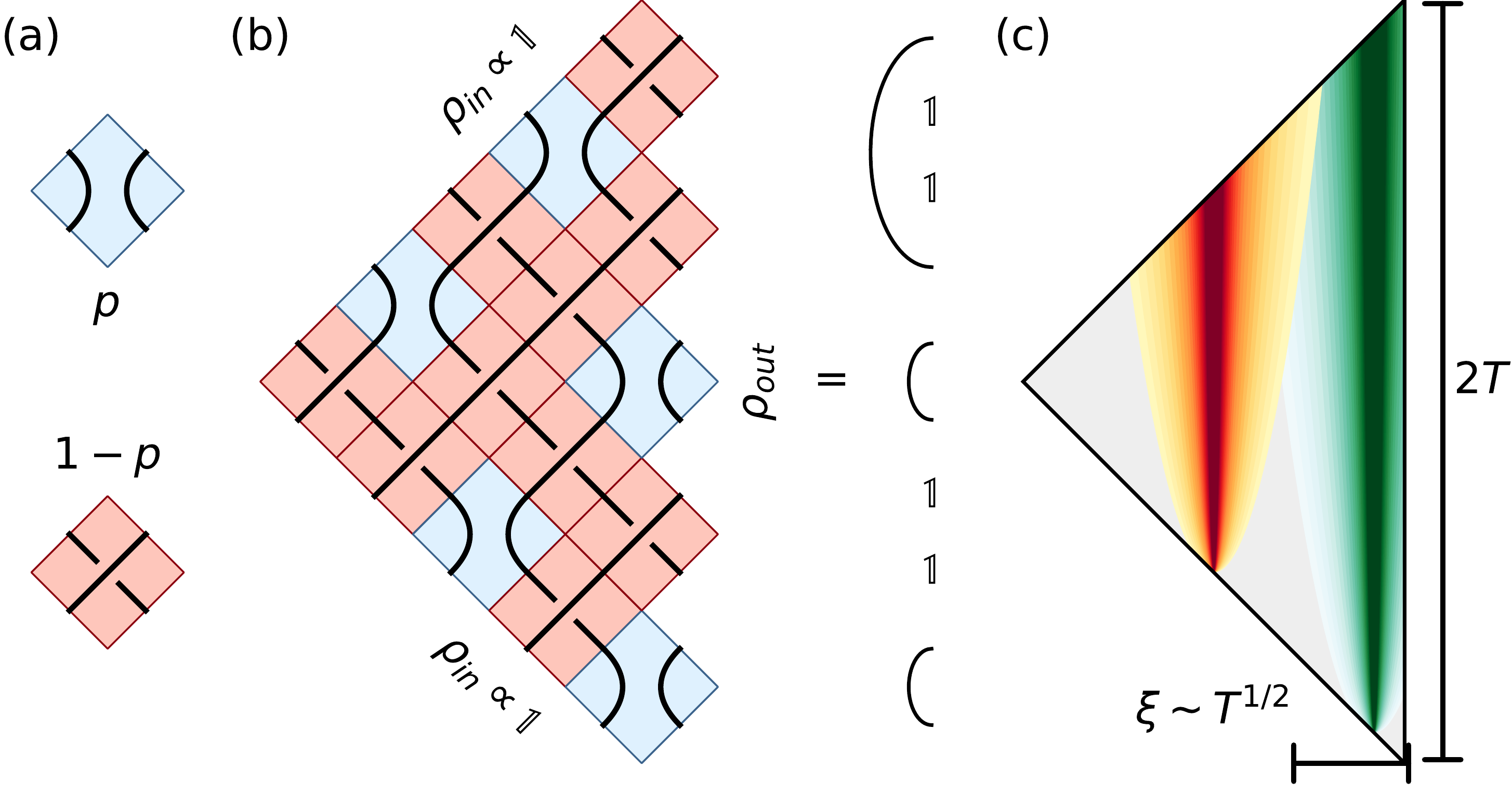}
    \caption{Critical phase of the non-interacting ($J=0$) Clifford circuit model. 
    (a) Allowed gates, $\mathbb 1$ (blue tile) and $\swap$ (red tile). 
    (b) A realization of the circuit for $T=5$. 
    The purification dynamics proceeds left to right;
    the input qubit worldlines ($\rhoin$, left) are fully mixed; the output state ($\rhoout$, right) contains Bell pairs (arcs) and fully mixed qubits ($\eye$ symbols). 
    (b) Coarse-grained view of the purification dynamics ($T\gg 1$). 
    Only qubit worldlines that enter within a distance $\xi\sim T^{1/2}$ of $\rhoout$ (e.g. green shaded parabola) are likely to diffuse to $\rhoout$ and contribute entropy.
    }
    \label{fig:loops}
\end{figure}

Setting $J=0$, the circuit maps to a loop model with two tiles, one associated to $\eye$ (probability $p$) and one to $\swap$ (probability $1-p$), see Fig.~\ref{fig:loops}(a);
the qubits move ballistically under $\swap$ gates and backscatter under $\eye$, thus tracing random walks with step size $\ell$ distributed exponentially, $\text{Prob}(\ell)\propto (1-p)^{|\ell|}$.
Worldlines that begin and end in $\rhoout$ define a pure Bell pair entirely contained in the system, and thus contribute no entropy;
on the contrary, wordlines that begin at the lightcone boundaries ($\rhoin$) and terminate in $\rhoout$, or viceversa, yield a fully mixed qubit in the output state and thus contribute one bit of entropy (Fig.~\ref{fig:loops}(b)).
How many such worldlines are there? 
Because the qubits undergo diffusion~\footnote{Because $\mathbb{E}(\ell)=0$ and $\mathbb{E}(\ell^2)$ is finite, this random walk gives rise to diffusion. The diffusion constant diverges as $p^{-2}$ for $p\to 0$.}, only those that enter the dynamics within $O(T^{1/2})$ steps of $\rhoout$ are likely to contribute entropy, see Fig.~\ref{fig:loops}(c). 
Thus we have $S_2(T) \sim \sqrt{T}$, and a stretched-exponential purification dynamics $\Tr(\rhoout^2) \sim e^{-c\sqrt{T}}$, to be compared with the exponential decay in the mixed phase.

{\it Subsystem purity and quantum code properties.}
Having established the existence (and prevalence) of the mixed phase in this class of models, it is interesting to investigate its properties, especially since the nature of the QECC defining the mixed phase remains in general poorly understood~\cite{Ippoliti2020, LiFisher2020, Fan2020}. 
To access such properties in experiment, one needs to measure not only the entropy of the entire state, but also of different subsystems. 

First we note that for \emph{contiguous}, \emph{even-sized} subsystems of the temporal slice where $\rhoout$ lives, $A = \{0\leq \tau<2t_A\}$, the subsystem purity obeys $\Tr (\rho_{\text{out},A}^2) = N(t_A)/N_\text{tot}$, and is therefore obtained for all $t_A\leq T$, \emph{at no additional cost}, by running the protocol up to time $T$. 
This follows from elision of all gates that lie outside a lightcone ending in $A$, see~\cite{Note1}.

To access general bipartitions, the protocol must be slightly modified.
The key idea is ``trace out'' qubits in the complement of the subsystem of interest, $\bar{A}$, by means of depolarization, e.g. by averaging over random unitary gates (see~\cite{Note1}).
We find that
\begin{equation}
    \Tr(\rho_{\text{out}, A}^2) = 2^{n_e-n_o} N_+(T; A)/N_\text{tot}\;,
    \label{eq:subsystem}
\end{equation}
where $n_e$ ($n_o$) is the number of even (odd) qubits in partition $\bar{A}$, on which a Bell pair is initialized (measured), and $N_+(T;A)$ is the number of successful runs based on a modified criterion:
the protocol \emph{cannot} fail on any of the $n_o$ Bell measurements in the partition $\bar{A}$; if a state other than $\ket{\bell}$ is obtained in such steps, it is reset to $\ket{\bell}$ and the protocol continues instead of failing~\cite{Note1}.

The purity for such non-contiguous bipartitions in stabilizer states can be used to obtain the contiguous code distance $d_\text{cont}$, an important property of the QECC that protects the mixed phase (see~\cite{Note1}). 
Numerically, we find that $\rhoout$ in the mixed phase defines a code with a power-law divergent, subextensive mutual information and contiguous distance (in the bulk of the system), similar to the phenomenology of the mixed phase in other Clifford models~\cite{Ippoliti2020, LiFisher2020}.


{\it Discussion.}
We have shown that a large class of non-unitary circuits allows direct experimental access to purification dynamics \emph{without postselection}, thus sidestepping a major obstacle toward the observation of entanglement phases in monitored circuits.
This is achieved by viewing the (non-unitary) spacetime duals of unitary gates as forced measurements.
Our protocol can be used to measure the purity of the whole system as well as arbitrary subsystems, and could enable the experimental investigation of the spatial entanglement structure and QECC properties in the mixed phase of these models. 

While the class of models we study is a measure-zero subset of all non-unitary circuits, it is nonetheless very large---in one-to-one correspondence with the space of unitary circuits made of nearest-neighbor two-qubit gates.
In this Letter we have studied a simple family of models as a demonstration;
a thorough exploration of this vast space and of the types of entanglement dynamics it may contain is a fascinating direction for future work.

\begin{acknowledgments}
\textit{Acknowledgments.} 
We thank Sarang Gopalakrishnan, Michael Gullans, David Huse, Xiaoliang Qi, Tibor Rakovszky and Dominic Williamson for valuable discussions.
This  work  was  supported  with  funding from  the  Defense  Advanced  Research  Projects  Agency (DARPA) via the DRINQS program (M.I.) and the US Department of Energy, Office of Science, Basic Energy Sciences, under Early Career Award No. DE-SC0021111 (V.K.). 
The views,  opinions and/or findings expressed are those of the authors and  should  not  be  interpreted  as  representing  the  official  views  or  policies  of  the  Department  of  Defense  or the  U.S.  Government.   
M.I.  was  funded  in  part  by  the Gordon and Betty Moore Foundation’s EPiQS Initiative through Grant GBMF4302 and  GBMF8686.
Numerical simulations were performed on Stanford Research Computing Center's Sherlock cluster.
\end{acknowledgments}

\bibliography{NoPostselection}

\end{document}


\title{Supplemental material: Postselection-free entanglement dynamics via spacetime duality}

\author{Matteo Ippoliti}
\affiliation{Department of Physics, Stanford University, Stanford, CA 94305, USA}
\author{Vedika Khemani}
\affiliation{Department of Physics, Stanford University, Stanford, CA 94305, USA}

\makeatletter
\renewcommand{\thesection}{S\arabic{section}}
\renewcommand{\theequation}{S\arabic{equation}}
\renewcommand{\thefigure}{S\arabic{figure}}

\maketitle 

\tableofcontents


\section{Details on spacetime duality}

Here we provide additional details on the notion of \emph{spacetime duality}, where one ``rotates'' the arrow of time by 90 degrees in spacetime to map a unitary gate $U$ to a generally different, non-unitary matrix $\tilde{U}$.

The duality transformation, $U_{i_1 i_2}^{o_1 o_2} = \tilde{U}_{i_1 o_1}^{i_2 o_2}$, rearranges the matrix entries as follows:
\begin{equation}    
U = 
    \begin{pmatrix}
    U_{00}^{00} & U_{00}^{01} & \red{U_{00}^{10}} & \red{U_{00}^{11}} \\
    \yellow{U_{01}^{00}} & \yellow{U_{01}^{01}} & U_{01}^{10} & U_{01}^{11} \\
    U_{10}^{00} & U_{10}^{01} & \green{U_{10}^{10}} & \green{U_{10}^{11}} \\
    \blue{U_{11}^{00}} & \blue{U_{11}^{01}} & U_{11}^{10} & U_{11}^{11}
    \end{pmatrix} 
    \qquad 
    \implies
    \qquad 
    \tilde{U} =
    \begin{pmatrix}
    U_{00}^{00} & U_{00}^{01} & \yellow{U_{01}^{00}} & \yellow{U_{01}^{01}} \\
    \red{U_{00}^{10}}  & \red{U_{00}^{11}} & U_{01}^{10} & U_{01}^{11} \\
    U_{10}^{00} & U_{10}^{01} & \blue{U_{11}^{00}} & \blue{U_{11}^{01}} \\
    \green{U_{10}^{10}}  & \green{U_{10}^{11}} & U_{11}^{10} & U_{11}^{11}
    \end{pmatrix} \;.
    \label{eq:duality_matrix}
\end{equation}
Black entries remain in their place, while two pairs of $1\times 2$ blocks (shown in color) are swapped.  

In the following we list a few examples. 
\begin{itemize}
    \item $U=\eye$:
$$
U = 
\begin{pmatrix}
1 & 0 & 0 & 0 \\
0 & 1 & 0 & 0 \\
0 & 0 & 1 & 0 \\
0 & 0 & 0 & 1
\end{pmatrix}
\implies \tilde{U} = \begin{pmatrix}
1 & 0 & 0 & 1 \\
0 & 0 & 0 & 0 \\
0 & 0 & 0 & 0 \\
1 & 0 & 0 & 1
\end{pmatrix}
= \frac{\eye + Z_1 Z_1}{2} (\eye + X_1 X_2)  = 2\ket{\bell}\bra{\bell},
\quad \ket{\bell} = (\ket{00} + \ket{11})/\sqrt{2}\;. $$
Thus $U=\eye$ dualizes to a forced Bell measurement, i.e. the projective measurement of $X_1X_2$ and $Z_1 Z_2$ with forced outcomes $(+1, +1)$.
\item $U = \textsf{CZ}$ (controlled-$Z$ gate):
$$
U = \begin{pmatrix} 1 & 0 & 0 & 0 \\ 0 & 1 & 0 & 0 \\ 0 & 0 & 1 & 0 \\ 0 & 0 & 0 & -1 \end{pmatrix}
\implies 
\begin{pmatrix}
1 & 0 & 0 & 1 \\
0 & 0 & 0 & 0 \\
0 & 0 & 0 & 0 \\
1 & 0 & 0 & -1
\end{pmatrix}
= \frac{\eye + Z_1 Z_2}{2} (Z_1 + X_1 X_2)
= \sqrt{2} \frac{\eye + Z_1 Z_2}{2} e^{-i \frac{\pi}{4} Y_1 X_2} Z_1 \;,
$$
i.e. the projective measurement of $Z_1 Z_2$ (with forced outcome $+1$) followed by a specific two-qubit unitary gate.
\item $U=\swap$:
$$
U = 
\begin{pmatrix}
1 & 0 & 0 & 0 \\
0 & 0 & 1 & 0 \\
0 & 1 & 0 & 0 \\
0 & 0 & 0 & 1
\end{pmatrix}
\implies 
\tilde{U} = 
\begin{pmatrix}
1 & 0 & 0 & 0 \\
0 & 0 & 1 & 0 \\
0 & 1 & 0 & 0 \\
0 & 0 & 0 & 1
\end{pmatrix} = U \;.
$$
Thus $\swap$ is a dual-unitary (and self-dual) gate. 
\end{itemize}

These examples illustrate that $\tilde{U}$ can generally involve a combination of forced measurement and unitary evolution. 
To characterize this more generally, we can perform a polar decomposition: 
$\tilde{U} = VH_0$, where $V$ is unitary and $H_0$ is a positive semidefinite matrix whose eigenvalues are the singular values of $\tilde{U}$, $H_0 \equiv \sqrt{\tilde{U}^\dagger \tilde{U}}$.
We note that the duality transformation merely reshuffles the matrix entries of $U$, hence it cannot change its Frobenius norm: $\| \tilde{U} \|^2 = \| U \|^2 = \Tr(U^\dagger U) = \Tr(\eye) = 4$.
Therefore $\Tr(H_0^2) = 4$. 
Redefining $H_0 \equiv H_0/2$ we have the identity mentioned in the main text,
\begin{equation}
    \tilde{U} = 2VH_0\;, 
    \qquad V \text{ unitary, }
    \qquad
    H_0\geq 0,
    \qquad \| H_0 \| = 1 \;.
\end{equation}
We note that $H_0$ can be seen as an element of a POVM set~\cite{NielsenChuangBook}, completed e.g. by $H_1\equiv \sqrt{\eye-H_0^2}$, i.e. another operator $H_1\geq 0$ such that $H_0^2 + H_1^2 = \eye$.
The set $\{H_0,H_1\}$ describes a binary measurement that yields one of two possible outcomes, $\alpha = 0,1$, with probabilities $p_\alpha = \Tr(\rho H_\alpha^2)$, while the (outcome-averaged) state evolves under a quantum channel $\rho\mapsto H_0 \rho H_0 + H_1 \rho H_1$.
If $H_{0,1}$ are orthogonal projectors then this is a projective measurement. More generally, this is a \emph{weak measurement}, achieved e.g. by coupling the system to an ancilla which is then projectively measured. 
The evolution described by the dualized gate $\tilde{U}$ is a \emph{forced} weak measurement, since the outcome is deterministically $\alpha = 0$. 

This gives rise to a generally non-trace-preserving evolution, $\rho\mapsto 4 V H_0 \rho H_0 V^\dagger$; however, the trace increase due to the prefactor and the trace loss due to forcing the $\alpha = 0$ outcome turn out to exactly compensate each other (with ``depolarizing'' or ``light-cone'' boundary conditions, discussed in the following), so that $\Tr \rhoout \equiv 1$ for hybrid circuits that are spatiotemporally dual to unitary circuits.
In particular this proves that the state cannot be annihilated by a forced measurement during the dynamics.


\section{Details on purity measurement protocol}

Here we provide a more detailed derivation of the purity measurement protocol, discuss the effect of ``depolarizing boundary conditions'', and how the protocol could be modified to implement conventional open boundary conditions.

\begin{figure}
    \centering
    \includegraphics[width=0.8\textwidth]{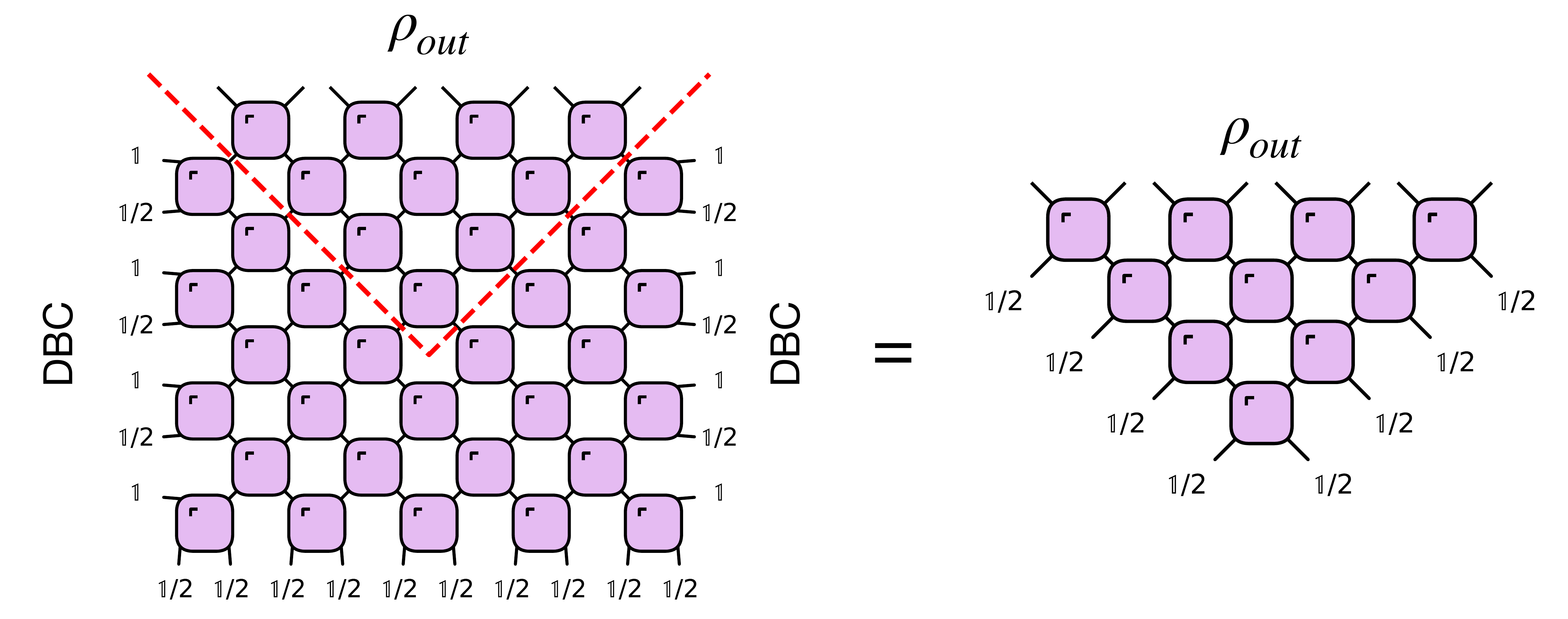}
    \caption{Purification dynamics with depolarizing boundary conditions (left). If the gates are unitary in the transverse direction, all gates outside the light cone (dashed line) are elided, and the dynamics reduces to light-cone boundary conditions (right). Purple boxes denote super-operators $u\otimes u^\ast$;  all tensor network legs are doubled, i.e. are legs of a density matrix.}
    \label{fig:dbc}
\end{figure}

\subsection{Depolarizing boundary conditions} 

The protocol we discuss is naturally formulated by starting from a fully mixed state (everywhere except on the central bond $\mathcal C$) and tracing out all final qubits (again everywhere except on $\mathcal{C}$). 
Under the spacetime duality transformation, these initial and final conditions map onto \emph{spatial} boundary conditions of a rather unusual type: the edge qubits are \emph{depolarized} at every time step, i.e. acted upon by a single-qubit fully depolarizing channel $\Phi(\rho_1) = \Tr(\rho_1) \frac{\eye}{2}$, where $\rho_1$ is a single-qubit density matrix, Fig.~\ref{fig:dbc}.
Thus for the whole state we have $\rho \mapsto \frac{\eye_1}{2} \otimes \Tr_{\{1,\tilde{L}\}}(\rho) \otimes \frac{\eye_{\tilde{L}}}{2}$.
This operation constantly injects some additional entropy into the system through the edges, and one may worry about its effect on the purification dynamics.

\begin{figure}
    \centering
    \includegraphics[width=0.5\textwidth]{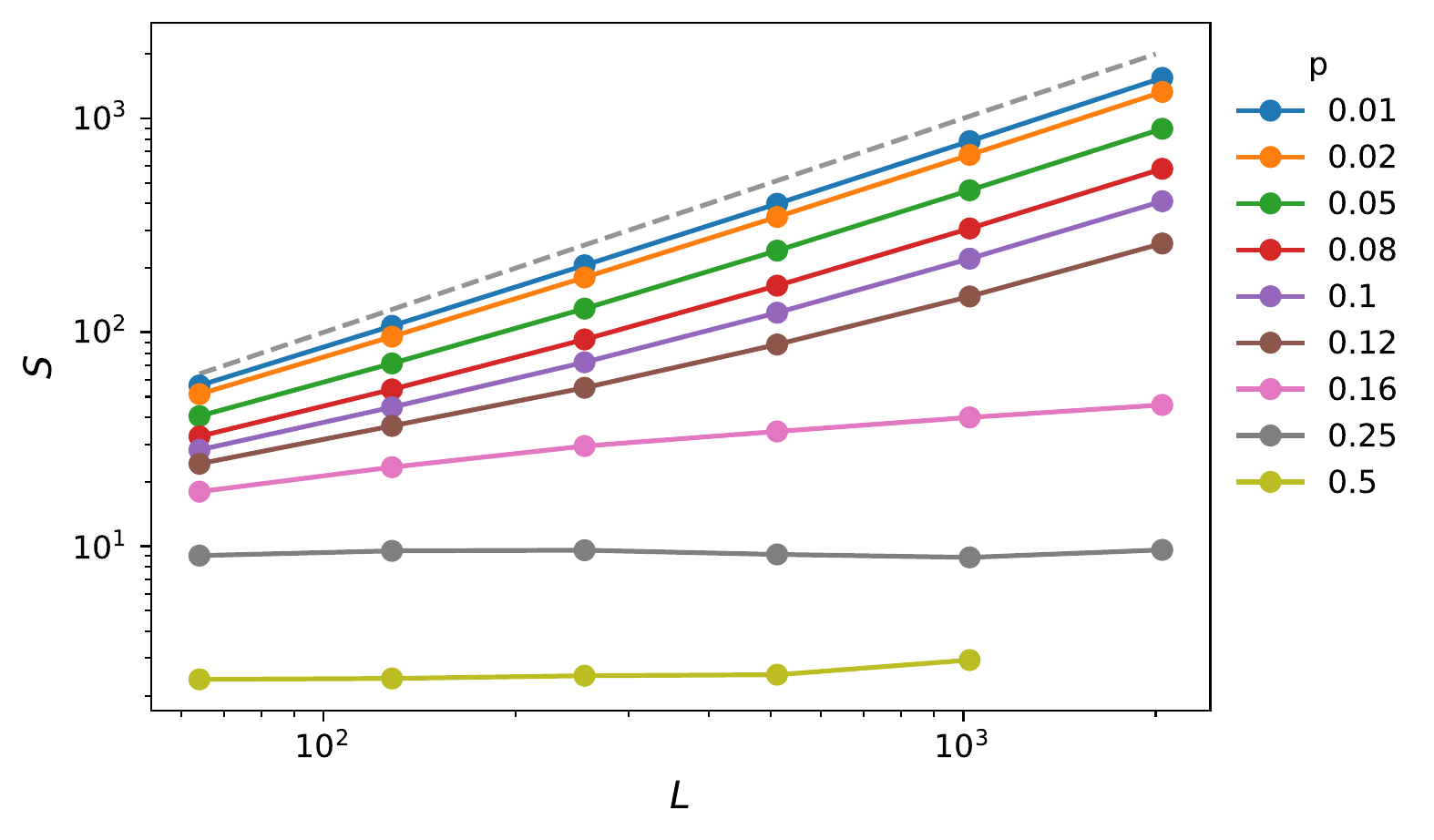}
    \caption{Numerical simulation of hybrid Clifford circuit model from Ref.~\cite{Li2018} with light cone boundary conditions: $L$ qubits $-L/2\leq x<L/2$ evolved for time $L/2$ with measurements taking place only inside the light cone $|x|<t$. 
    $p$ is the measurement rate; $p_c=0.16$ is the known location of the critical point, separating a mixed phase ($p<p_c$) from a pure phase ($p>p_c$).
    With these boundary conditions, the pure phase has an $O(1)$, rather than vanishing, amount of entropy.
    }
    \label{fig:wedgeLCF}
\end{figure}

We performed numerical simulations of the original Clifford unitary-projective model in Refs.~\cite{Li2018, Gullans2020PRX}, featuring arbitrary gates and projective $Z$ measurements, starting from a maximally mixed state and depolarizing the edge qubits at every time step.
We found that the purification phase diagram is unchanged, see Fig.~\ref{fig:wedgeLCF}; namely we find that (i) the value of the critical measurement rate $p_c\simeq 0.16$ is unchanged, (ii) the $p$-dependent entropy density in the mixed phase is quantitatively unchanged, and (iii) the only difference is that the pure phase does not achieve zero entropy, but rather a finite, $p$-dependent amount of entropy (it still achieves zero entropy density as $L\to\infty$).
This is reasonable, as the entropy constantly injected through the sides is purified in $O(1)$ time, but not instantly.

Within the class of hybrid circuits that are dual to unitary circuits, the depolarizing boundary conditions described above have a global effect on the circuit.
As is clear from viewing the circuit in the unitary direction (arrow of time $t$), all gates that are not in the intersection of the future light cone of the initial Bell state $|P_{\mathcal C})$ and the past light cone of the final Bell measurement $(P_{\mathcal C}|$ are elided (Fig.~\ref{fig:dbc}).
In the non-unitary time direction (arrow of time $\tilde{t}$), this amounts to \emph{only allowing measurements inside a light cone} $|\tilde{x}|<\tilde{t}$, so that the state outside such light cone remains fully mixed. 
This places a constraint on the maximum duration of the hybrid time evolution: given a fixed ``size'' $\tilde{L} = 2T$ for the output state $\rhoout$, measurements (and thus purification) can begin \emph{at most} $\tilde{T} = T = \tilde{L}/2$ time steps prior. 
Thus for instance it is not possible to use this approach to probe the very late-time purification dynamics at $\tilde{T} \gg \tilde{L}$ in the mixed phase~\cite{Fidkowski2020}.

\subsection{Normalization}
In the main text we claim that
\begin{equation}
    \Tr(\rhoout^2|_{\tilde{L}=2T, \tilde{T}=T}) = N_+(T)/N_\text{tot} \;,
    \label{eq:supp_protocol}
\end{equation} however we only prove the relationship up to prefactors.
Here we show that Eq.~\eqref{eq:supp_protocol} holds as written, i.e. with a proportionality constant of 1.

\subsubsection{Trace of $\rhoout$}

\begin{figure}
    \centering
    \includegraphics[width=\textwidth]{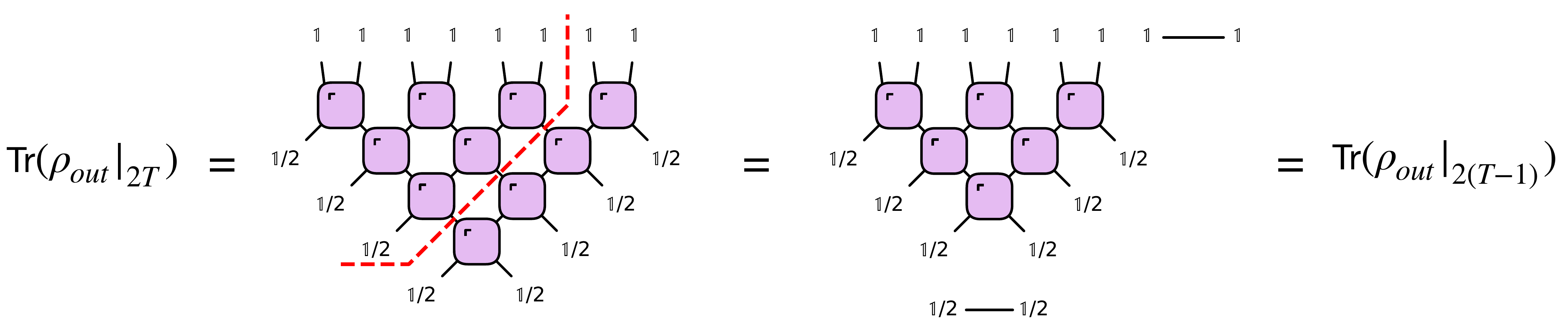}
    \caption{Normalization of $\rhoout$. Using unitarity of the gates in the transverse direction, all gates outside the dashed line can be elided, and the trace of $\rhoout$ on $\tilde{L} = 2T$ qubits (left) can be reduced to the trace of $\rhoout$ on $\tilde{L}' = 2(T-1)$ qubits (right), with a prefactor of $\Tr(\mathbb 1) \Tr(\mathbb{1}/4) = 1$. Iterating the argument yields $\Tr(\rhoout)=1\, \forall T$.}
    \label{fig:trace}
\end{figure}

We begin by observing that, despite the non-trace-preserving nature of the gates $\tilde{u}$, the state $\rhoout$ is in fact normalized, $\Tr(\rhoout)\equiv 1$, with `light cone' boundary conditions as discussed above. 
The argument is illustrated in Fig.~\ref{fig:trace} and follows naturally from unitarity of the gates $u$ that make up the evolution in the transverse direction: whenever a gate is contracted with single-qubit identity matrices on \emph{both} its right (or left) legs, it can be elided (a consequence of unitarity, $u (\eye \otimes \eye) u^\dagger = uu^\dagger = \eye \otimes \eye$).
This creates a new `corner' in the tensor network, and the reasoning can be iterated to elide all gates one by one, leaving a circuit-independent constant. Direct inspection yields $\Tr(\rhoout) = \Tr(\rhoin) = 1$.

\subsubsection{Normalization of purity}

\begin{figure}
    \centering
    \includegraphics[width=\textwidth]{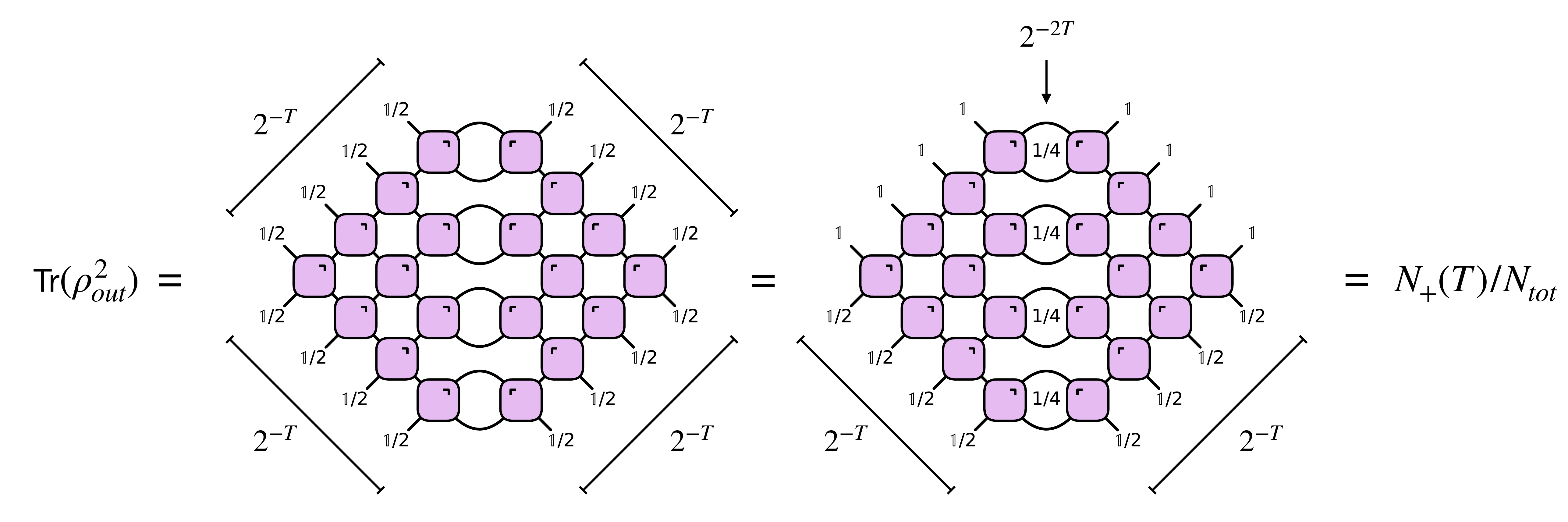}
    \caption{Fixing the proportionality constant between $\Tr(\rhoout^2)$ and the protocol's success probability $N_+(T)/N_\text{tot}$. Both quantities are equal to $2^{-4T}$ times the same tensor network contraction. Left: each copy of $\rhoin$ comes with a normalization of $2^{-2T}$. Right: in the measurement protocol, the initial state carries a normalization of $2^{-2T}$, while the iterated Bell-state projections on the central bond carry an additional normalization of $2^{-2T}$ (from the factors of $\sqrt{2}$ in $\ket{\bell}$).}
    \label{fig:purity}
\end{figure}

Here we provide an explicit derivation of the normalizations on the two sides of Eq.~\eqref{eq:supp_protocol}.
Let $\mathcal{D}(T)$ denote the tensor network in Fig.~\ref{fig:purity} with single-qubit identity matrices contracted on all dangling legs, \emph{without any prefactors}.
Focusing first on the hybrid time direction, we have $\Tr(\rhoout^2) = 2^{-4T}\mathcal{D}(T)$, 
to account for the $2T$ incoming qubits in each copy of $\rhoin$, each of which carries a normalizing factor of $1/2$.
Next we focus on the unitary time direction. Here we have $2T$ factors of $1/2$ from the fully mixed input qubits in the two sides of the system, $\mathcal{L}$ and $\mathcal{R}$; 
additionally, each instance of the projector $P_{\mathcal C}$ carries a factor of $1/2$ (i.e., the horizontal leg contractions at the central bond $\mathcal C$ are equal to $2P_{\mathcal C}$).
Since there are $2T$ such projectors at $\mathcal C$ over the course of the protocol ($T$ initializations and $T$ measurements), we have overall $N_+(T)/N_\text{tot} = 2^{-4T} \mathcal{D}(T)$.
Thus both sides of Eq.~\eqref{eq:supp_protocol} are equal to $2^{-4T} \mathcal{D}(T)$, and the proportionality constant is 1.

\subsection{Open boundary conditions}

\begin{figure}
    \centering
    \includegraphics[width=0.7\textwidth]{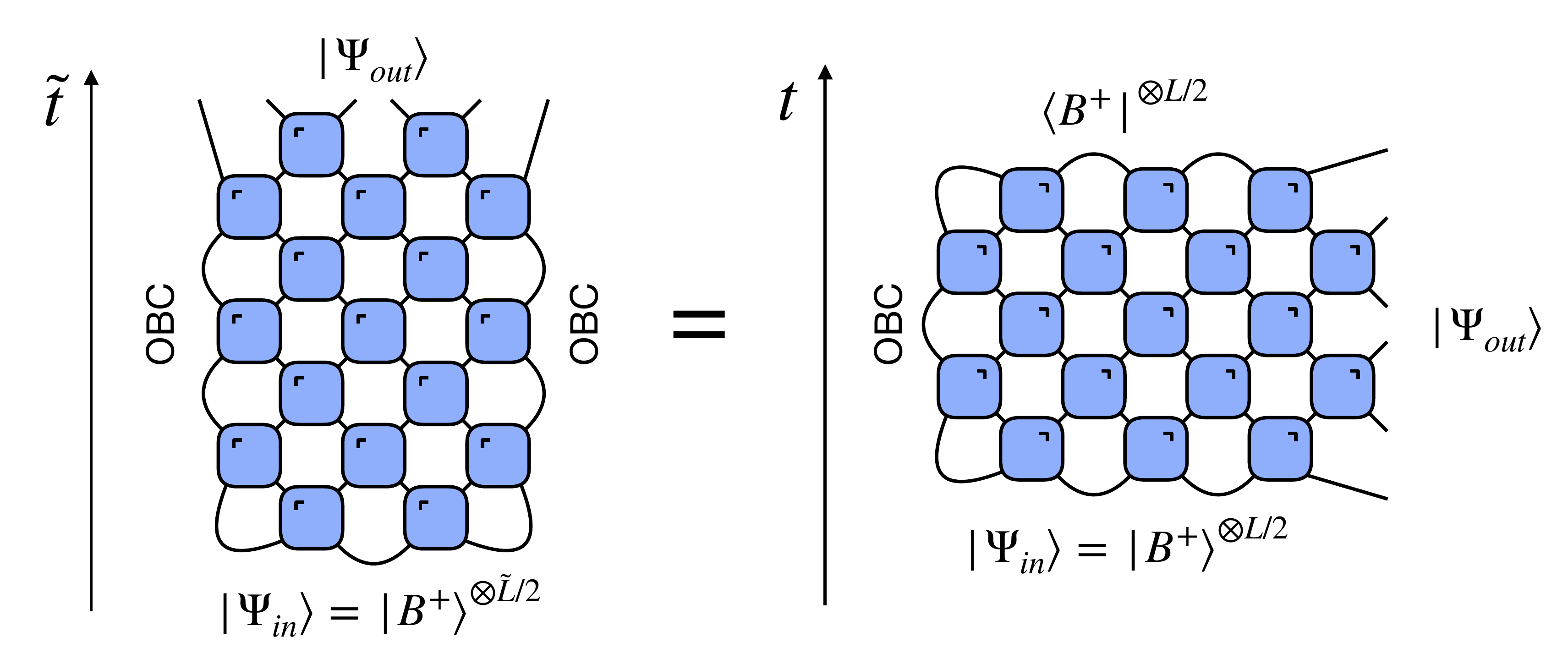}
    \caption{Left: pure-state hybrid circuit evolution. Blue boxes represent single unitary gates $u$ (as opposed to the purple boxes in Figs.~\ref{fig:dbc}, \ref{fig:trace}, \ref{fig:purity} representing super-operators $u\otimes u^\ast$). An initial product state on two-qubit subsystems $\ket{\Psi_\text{in}} = \ket{\bell}^{\otimes \tilde{L}/2}$ ($\tilde{L}=6$ in the figure) is evolved via a non-unitary brickwork circuit, with open boundary conditions, producing a (pure) output state $\ket{\Psi_\text{out}}$.
    Right: under spacetime duality, the same state is obtained on a temporal slice at the right edge of the system. A pure product state $\ket{\Psi_\text{in}} = \ket{\bell}^{\otimes L/2}$ ($L=6$ in the figure) is evolved with a unitary brickwork circuit, with open boundary conditions on the left edge (the dangling legs at the right edge could be seen as qubits swapped in and out of the system, see Sec.~\ref{sec:prep}); at the final time step, the $L$ ``system'' qubits are projected onto the Bell-pair initial state $\bra{\Psi_\text{in}}$ (with postselection). 
    }
    \label{fig:purestate}
\end{figure}

One could replace the depolarizing boundary conditions with more conventional open boundary conditions (OBCs) at the expense of postselecting on $L/2$ Bell measurement outcomes.
This can be achieved by having an initial state (in the ``laboratory'' time direction) made of $L/2$ nearest-neighbor Bell pairs, $\ket{\Psi_\text{in}} = \ket{\bell}^{\otimes L/2}$, where $L$ is now unrelated to the protocol duration $T$ (i.e. it can be longer or shorter), and projecting onto the same state after $T$ time steps (which requires postselection), as shown in Fig.~\ref{fig:purestate}.

Viewed in the non-unitary time direction, the Bell-pair dimerization of the initial and final (postselected) states represents hard-wall boundary conditions on the left and right boundaries, see Fig.~\ref{fig:purestate} (left).
The initial state of the hybrid dynamics can likewise be chosen to be pure, e.g. by having open boundaries at the edges of the system in the laboratory.
This would allow one to simulate \emph{entanglement dynamics}, rather than the purification dynamics this work focuses on. In addition, OBCs remove the light-cone constraint and allow arbitrary independent values of $\tilde{L} \equiv 2T$ and $\tilde{T}\equiv L/2$. 
The postselection cost is $2^L$ ($L/2$ Bell measurements with 4 possible oucomes each); this is exponentially lower than the $e^{O(LT)}$ postselection cost for conventional simulation of unitary-projective models with finite measurement density in spacetime, assuming a deep circuit with $T = O(L)$.


\section{State preparation protocol \label{sec:prep}}

In the main text we have described a method to evaluate the purity of the hybrid circuit output $\rhoout$ `on the fly', without ever storing the state. 
However, it may be useful for certain applications to actually prepare the state on a spatial slice of the circuit. Here we describe how to do so with modest overhead ($O(T)$ ancillas and $\swap$ gates). 

\begin{figure}
    \centering
    \includegraphics[width=0.4\textwidth]{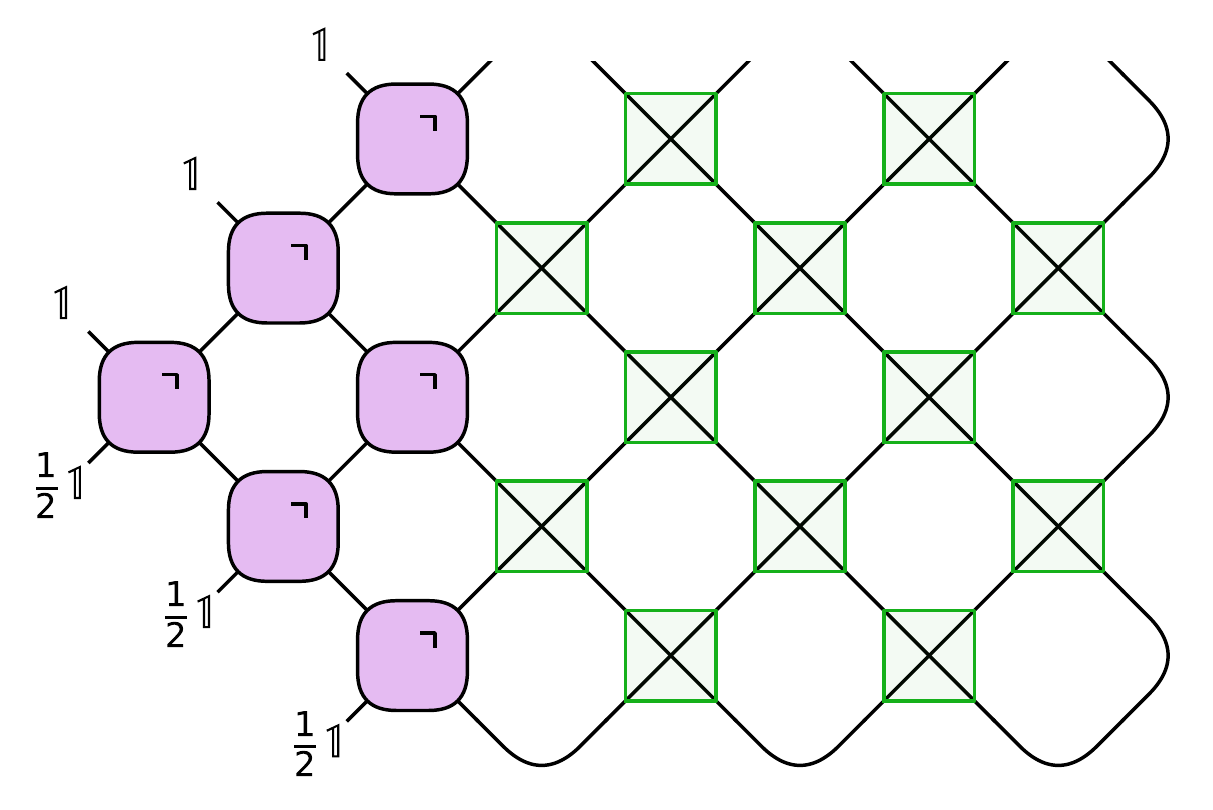}
    \caption{State preparation protocol. The purple boxes represent two-qubit super-operators $u\otimes u^\ast$ and all qubit worldlines should be seen as doubled (i.e. as legs of a density matrix).
    Fully mixed qubit states $\frac{1}{2}\eye$ (bottom left) are evolved by the $u$ gates and eventually traced out (top left).
    Ancillae initialized in Bell pair states (bottom right) are evolved with $\swap$ gates (transparent green boxes), with open boundary conditions on the right. 
    As a result, the temporal slice where $\rhoout$ lives is ``teleported'' onto a spatial slice in the ancillary system (top right).}
    \label{fig:prep}
\end{figure}

The idea is to prepare $2T$ ancillary qubits on which to store the output state $\rhoout|_{2T}$, and to use a version of `gate teleportation'~\cite{NielsenChuangBook} in order to transfer the state from the temporal slice where it is naturally generated onto the ancillas.
In practice one needs a chain of $3T$ qubits: the $T$ qubits on the left are initialized in a fully mixed state, while the $2T$ qubits on the right are initialized in nearest-neighbor Bell pairs $\ket{\bell}^{\otimes T}$, see Fig.~\ref{fig:prep}.
Then the left qubits are evolved via the circuit $U_M$, while the right qubits (ancillas) are evolved by a $\swap$ circuit with open boundary conditions on the right wall. 
As can be seen in Fig.~\ref{fig:prep}, the qubit wordlines are geometrically reflected by the Bell-pair initial condition and by the hard-wall boundary on the right, and thus end up in the correct order on a spatial slice at time $T$.


\section{Details on subsystem purity measurement}

Here we discuss in more detail how to measure the purity of arbitrary subsystems $A\subseteq \{0,1,\dots 2T-1\}$, $\Tr(\rho_{\text{out},A}^2)$ with $\rho_{\text{out},A} = \Tr_{\bar A}\rhoout$.

\subsection{Contiguous subsystems \label{sec:subsystem_cont}}

First we consider contiguous, even-sized bipartitions, $A = \{0\leq \tau<2t_A\}$, $\bar{A} = \{2t_A\leq \tau <2T\}$. 
We observe that tracing out a subsystem $\bar{A}$ near an edge in $\rhoout$, combined with the lightcone boundary conditions, leads to the elision of all gates outside a lightcone terminating in $A$ (by the same mechanism as Fig.~\ref{fig:trace}).
The remaining gates define a state $\rhoout$ with the same lightcone boundary conditions, but on a smaller number of qubits, $\tilde{L}' = 2t_A$, i.e., 
\begin{equation}
\Tr_{\bar{A}} \left( \rhoout|_{\tilde{L}=2T, \tilde{T}=T}\right) 
= \rhoout|_{\tilde{L}'=2t_A, \tilde{T}'=t_A} \;.
\end{equation}
Thus the purity of $\rho_{\text{out},A}$ is obtained from the full-system purity measurement at time $t_A$:
\begin{equation}
\Tr(\rho_{\text{out},A}^2) = N_+(t_A)/N_\text{tot} \;.
\label{eq:supp_contiguous}
\end{equation}
Thus by running the purity measurement protocol up to time $T$ and storing the number of successful runs at all intermediate times, $\{N_+(t_A): 0<t_A\leq T \}$, one gets not only the purity of the whole state $\rhoout$, but also of all contiguous, even-sized subsystems $A = \{0\leq \tau < 2t_A\}$, at no additional cost.

Using these data, one can obtain the linear combination of entropies
\begin{equation}
I_2^{(\alpha)} (A:\bar{A}) = S_\alpha(A) + S_\alpha(\bar{A}) - S_\alpha (A\cup \bar{A})
\label{eq:pseudo_I2}
\end{equation}
for Renyi index $\alpha=2$, as
\begin{equation}
I_2^{(2)} = \log \frac{N_+(T)}{N_+(t_A) N_+(T-t_A)} \;.
\end{equation}
For stabilizer states, entropies do not depend on $\alpha$ and thus $I_2^{(2)}(A:\bar{A}) = I_2^{(1)}(A:\bar{A})$ ($\alpha=1$ being the Von Neumann entropy), which is the mutual information between $A$ and its complement. 
In general though $I_2^{(2)}$ is not guaranteed to be positive and is thus not a mutual information.

\subsection{General subsystems}

Here we consider arbitrary subsystems $A$ of $\rhoout$.
We prove the following statement from the main text,
\begin{equation}
    \Tr(\rho_{\text{out},A}^2) = 2^{n_e-n_o} N_+(T;A)/N_\text{tot} \;,
    \label{eq:supp_subsystem}
\end{equation}
and clarify the definition of $n_{e/o}$ and of $N_+(T;A)$.

The problem with computing subsystem purities in our setting is how to perform the partial trace \emph{during the dynamics}, as $\rhoout$ exists on a time-like slice of the circuit.
One option is to use the state preparation protocol in Sec.~\ref{sec:prep}, repeatedly prepare identical copies of the state, and use a technique based on randomized measurements to obtain the entropy of mixed states~\cite{Elben2020}.
However, this is not necessary: a small modification to the protocol for full-system purity allows measurement of subsystem purities `on the fly', without having to prepare the satate in space.

\begin{figure}
    \centering
    \includegraphics[width=0.8\textwidth]{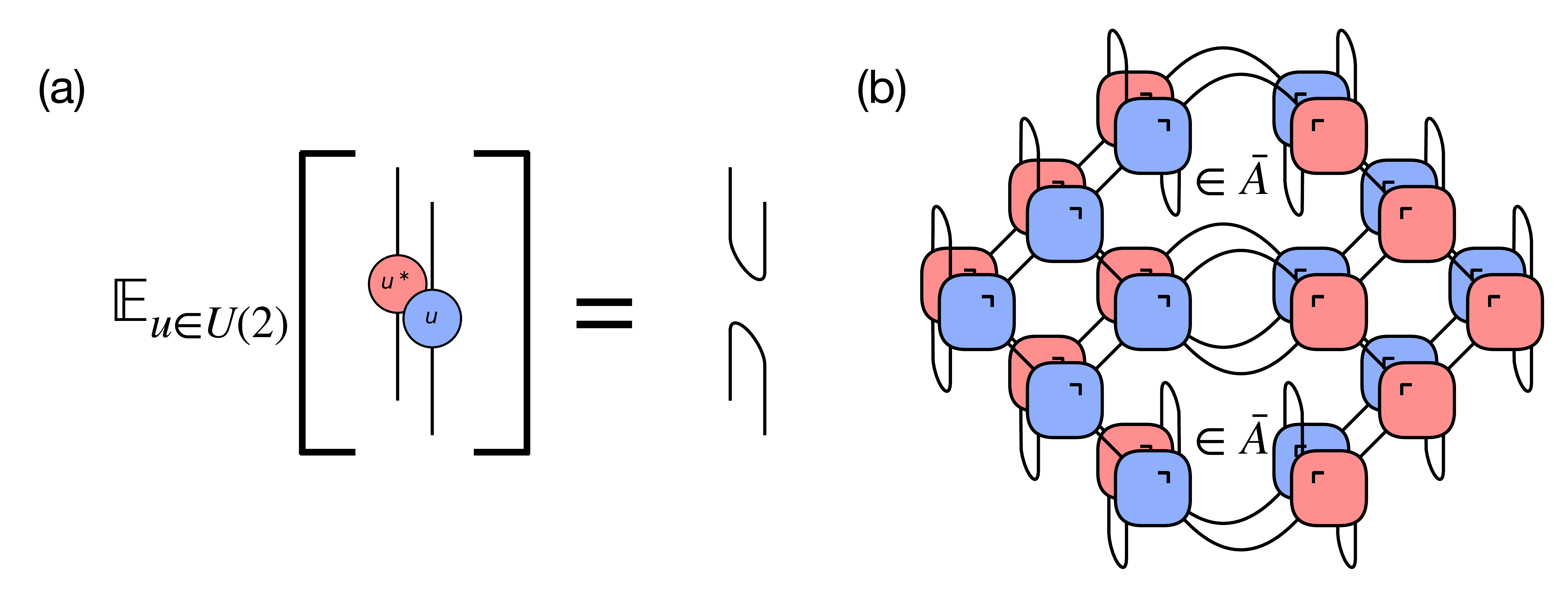}
    \caption{(a) Depolarizing a qubit by averagin random gates.
    (b) Protocol for subsystem purity measurement, $T=3$ (cf. Fig.~2(d) main text). This represents the purity of subsystem $A=\{0, 2, 3, 5 \}$, obtained by tracing out the complement $\bar{A} = \{1, 4 \}$. }
    \label{fig:depolarizing}
\end{figure}

The key idea is to \emph{depolarize} any qubits that belong to $\bar{A}$ (and thus need to be traced out) by acting on them with random, uncorrelated single-qubit gates, and to average the results of the protocol over many independent realizations.
This uses the property of Haar-random averages, 
\begin{equation}
    \mathbb{E}_{u\in U(2)} (u_i \rho u_i^\dagger) = \Tr_i(\rho) \otimes \frac{\eye_i}{2} \;.
    \label{eq:depolarization}
\end{equation}
The Haar average can in fact be simplified to a random Clifford average, by exploiting the 1-design property. 
The result of averaging many runs of this pure-state dynamics, $\ket{\psi}\mapsto u \ket{\psi}$, is to implement a perfect single-qubit depolarizing channel $\Phi(\rho) \mapsto \Tr(\rho) \eye/2$, see Fig.~\ref{fig:depolarizing}(a).
This implements the `tracing out' one needs to compute the reduced density matrix, Fig.~\ref{fig:depolarizing}(b).

Because $\rhoout$ lives in a temporal slice of the circuit, the subsystems $A$ and $\bar{A}$ are themselves parts of this temporal slice, i.e. $A \subseteq \{0,\dots 2T-1\}$.
Qubits in this time-like slice can play two different roles: even qubits $\mathcal{E} \equiv \{0,2,\dots 2T-2\}$ at time steps where a Bell projector $P_{\mathcal C}$ is initialized; and odd qubits $\mathcal{O} \equiv \{1,3,\dots 2T-1\}$ at time steps where a Bell projector $P_{\mathcal C}$ is measured, see Fig.~\ref{fig:normalizations}. It is helpful to analyze these separately.

\subsubsection{Depolarizing even qubits}
\begin{figure}
    \centering
    \includegraphics[width=\textwidth]{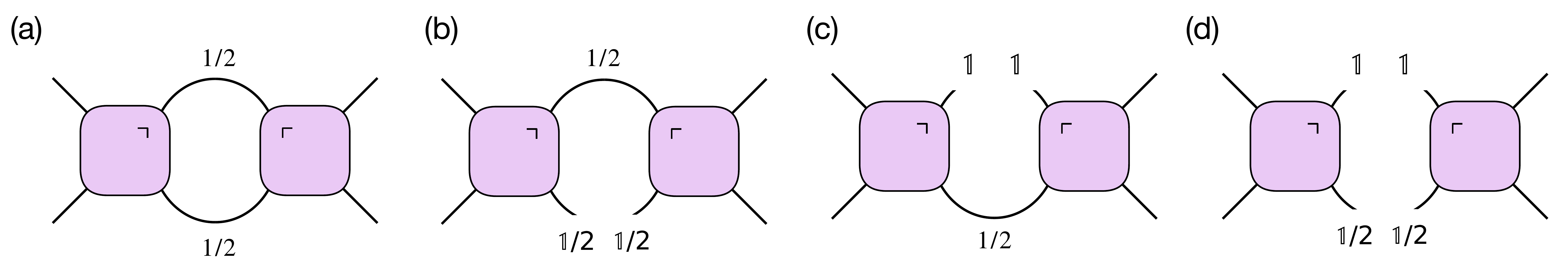}
    \caption{Accounting of normalizations for subsystem purity calculation.
    One timestep for the central pair of qubits (cf. Fig.~\ref{fig:purity}). 
    (a) No qubits are depolarized. Preparing $|P)$ and measuring $(P|$ implements the desired diagram up to \emph{two} factors of $1/2$.
    (b) Even qubit is depolarized.
    Preparing $\frac{1}{4} |\eye\otimes \eye)$ and measuring $(P|$ implements the desired diagram up to \emph{three} factors of $1/2$. 
    (c) Odd qubit is depolarized.
    Preparing $|P)$ and tracing out region $\mathcal{C}$ implements the desired diagram up to \emph{one} factor of $1/2$.  
    (d) Both the even and the odd qubits are depolarized. 
    Preparing $\frac{1}{4} |\eye\otimes \eye)$ and tracing out region $\mathcal{C}$ implements the desired diagram up to \emph{two} factors of $1/2$.
    Overall, each instance of (b) produces an excess factor of $1/2$, while each instance of (c) produces an excess factor of 2, thus the normalization $2^{n_e-n_o}$ in Eq.~\eqref{eq:supp_subsystem}.
    }
    \label{fig:normalizations}
\end{figure}

Let us consider a qubit $\tau \in \mathcal E$ in $\rhoout$, i.e. a time step $\tau$ where a Bell state $\ket{\bell}$ has just been created.
To depolarize this qubit, one would apply a random Clifford gate on, for example, qubit $j=-1$ (either one of $\pm 1 \in \mathcal{C}$ would be equally good) at time step $\tau$:
$$
\mathbb E_{u}\left[ u_{-1} \otimes \eye_{1} \ket{\bell}\bra{\bell} u_{-1}^\dagger \otimes \eye_1\right] 
= \frac{\eye_{-1}}{2} \otimes \Tr_{-1}(\ket{\bell}\bra{\bell}) = \frac{1}{4} \eye_{\mathcal C} \;.
$$
(The state on the rest of the system outside $\mathcal{C}$ is omitted.)
The outcome of this is a fully mixed state on $\mathcal C$. The protocol can then continue like in the normal (full-system purity) case.
However, there is an extra normalization to take into account: the state $\eye_{\mathcal C}$ carries a normalizing factor of $1/4$ relative to the tensor network diagram $\mathcal{D}$ it aims to represent, Fig.~\ref{fig:normalizations}(b). On qubits that belong to $A$ (and are thus not depolarized), the Bell state $\ket{\bell}\bra{\bell}$ carries only a factor of $1/2$ relative to the tensor network $\mathcal{D}$ (Fig.~\ref{fig:normalizations}(a)). 
These excess factors of $1/2$, one for each qubit in $\mathcal E \cap \bar{A}$, must be compensated by a factor of $2^{n_e}$, with $n_e\equiv |\mathcal E \cup \bar{A}|$. 
This explains part of the normalization in Eq.~\eqref{eq:supp_subsystem}.

\subsubsection{Depolarizing odd qubits}
Next we consider a qubit $\tau \in \mathcal{O}$ in $\rhoout$, i.e. a time step $\tau$ where a Bell measurement is about to take place.
In this case it is sufficient to perform the Bell measurement and `reset' the state to $\ket{\bell}$ regardless of the measurement outcome.
This is done by acting with a single-qubit gate, on e.g. qubit $j=1$, conditioned on the Bell measurement outcome as follows:
$$
\ket{\bell} = \frac{\ket{00}+\ket{11}}{\sqrt{2}} \mapsto \eye_{1};
\qquad
\frac{ \ket{00}-\ket{11}}{ \sqrt{2} } \mapsto Z_1;
\qquad 
\frac{\ket{01}+\ket{10}}{\sqrt{2}} \mapsto X_1;
\qquad 
\frac{ \ket{01}-\ket{10} }{ \sqrt{2}} \mapsto Y_1.
$$
Importantly none of the outcomes determine a `failure' -- the protocol continues regardless. This modified `success' criterion defines the numerator $N_+(T;A)$ in Eq.~\eqref{eq:supp_subsystem}. 
The process we described corresponds to 
$$
\rho \mapsto P_{\mathcal C} \otimes \Tr_{\mathcal C}(\rho)
$$
i.e. the preexisting state of qubits $\pm 1\in \mathcal C$ is simply discarded and replaced by the new state. 
This process exactly implements the contraction in the tensor network $\mathcal D$, without normalizations (Fig.~\ref{fig:normalizations}(c)); in comparison, on qubits that belong to $A$ the state transforms as $\rho \mapsto P_{\mathcal C} \otimes \Tr_{\mathcal C} (\rho P_{\mathcal C})$ which carries a factor of $1/2$ relative to the tensor network $\mathcal{D}$. 
The excess factors of $2$, one for each qubit in $\mathcal{O} \cap \bar{A}$, must be compensated by a factor of $2^{-n_o}$, with $n_o\equiv |\mathcal O \cap \bar{A}|$.
This explains the remaining part of the normalization in Eq.~\eqref{eq:supp_subsystem}

\subsubsection{Sanity check}

We can verify the correctness of Eq.~\eqref{eq:supp_subsystem} by checking its consistency with (i) the contiguous-subsystem result Eq.~\eqref{eq:supp_contiguous}, and (ii) the result for a simple circuit.

\begin{enumerate}[label=(\roman*)]
\item
A bipartition $A = \{0\leq \tau < 2t_A\}$, $\bar{A} = \{2t_A\leq \tau < 2T\}$ has $n_e = n_o = |\bar{A}|/2 = T-t_A$, thus $2^{n_e-n_o}= 1$; moreover, since the protocol can never fail after the $t_A$-th measurement, we have $N_+(T;A) = N_+(t_A)$. Plugging these into Eq.~\eqref{eq:supp_subsystem} yields Eq.~\eqref{eq:supp_contiguous}.
\item
Take $U_M = \eye$ (the all-identity circuit). Its (non-unitary) spacetime-dual $M$ consists of Bell measurements arranged in a brickwork pattern; its output state is trivially $\rhoout \propto P^{\otimes T}$, a pure state made of Bell pairs on all even (or odd) bonds. 
This state has maximal entanglement entropy $S_2 = T$ for a non-contiguous bipartition consisting of $\mathcal{E}$ and $\mathcal{O}$ (even vs odd qubits), since all $T$ Bell pairs straddle the cut.
We can compute this in two ways -- setting $A=\mathcal{E}$ and $\bar{A} = \mathcal{O}$ or vice versa -- and expect the same result since $\rhoout$ is pure.
\begin{itemize}
\item ($A=\mathcal{O}$, $\bar{A}=\mathcal{E}$)
The central bond is sequentially depolarized and measured in the Bell basis. The outcome of each measurement is fully random, with $1/4$ success probabiliy; thus $N_+(T;A)/N_\text{tot} = 4^{-T}$. However, because $n_e = |\bar{A}\cap \mathcal E| = |\mathcal E| = T$, the normalization yields $\Tr \rho_{\text{out},A}^2 = 2^T \cdot 4^{-T} = 2^{-T}$, i.e. $S_2 = T$, as expected.
\item ($A = \mathcal{E}$, $\bar{A}=\mathcal{O}$) 
The central bond is sequentially prepared in the Bell state $P = \ket{\bell}\bra{\bell}$ and measured in the Bell basis, which trivially always yields the outcome $\ket{\bell}$.
Thus $N_+(T;A)/N_\text{tot} = 1$ (the protocol can never fail), while the normalization is $2^{-n_o} = 2^{-T}$, since $n_o = |\bar{A} \cap \mathcal{O}| = |\mathcal{O}| = T$.
Again, we find $S_2 = T$ as expected.
\end{itemize}
\end{enumerate}

\section{Quantum code properties in the mixed phase}

Here we report results of numerical simulations of the QECC properties (stabilizer length distribution, mutual information, and contiguous distance) for the Clifford model in the mixed phase, $0<p<1$ and $J>0$.

\begin{figure}
    \centering
    \includegraphics[width=0.9\textwidth]{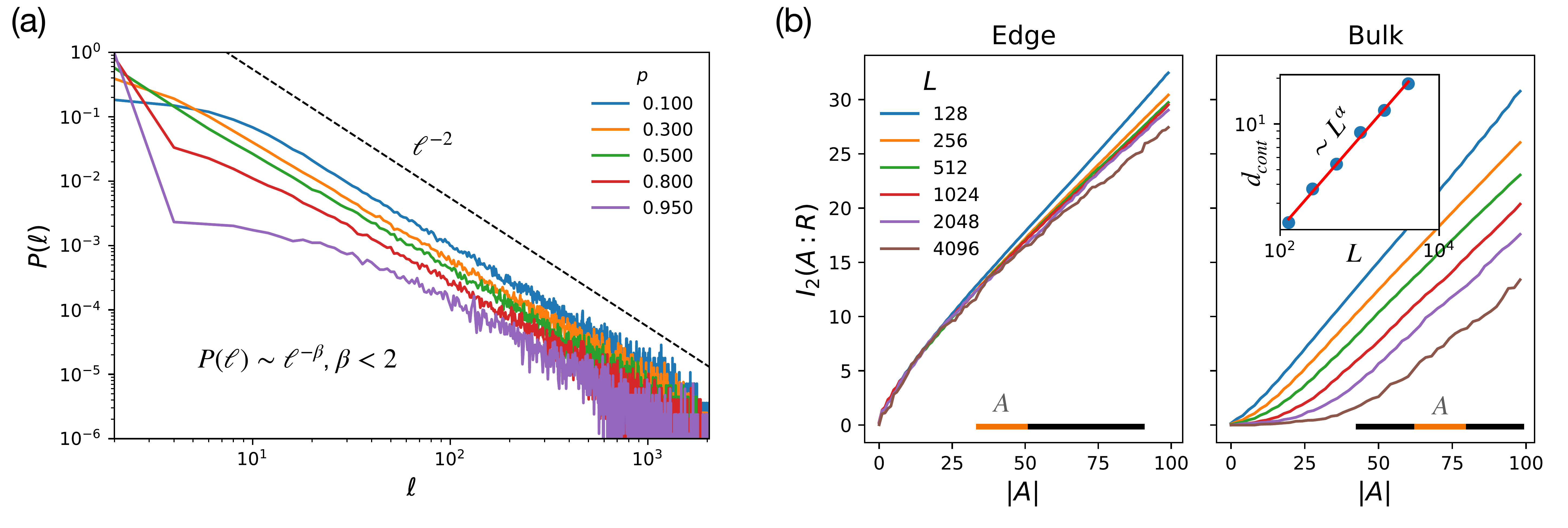}
    \caption{QECC properties in the mixed phase: numerical results from Clifford simulations of up to $L=4096$ qubits wiht `light cone' boundary conditions.
    (a) Stabilizer length distribution $P(\ell)$ for a system of size $L=2048$, $J=1$, variable $p$.
    $P(\ell)$ has a power-law tail with exponent $\beta < 2$, indicating a power-law, rather than logarithmic, divergence of the mutual information $I_2(A:\bar{A}) \sim |A|^{2-\beta}$ (we find $2-\beta\simeq 0.2$).
    (b) Mutual informtion $I_2(A:R)$ between the reference qubits $R$ and contiguous subsystems $A$ near the edge (left) and in the bulk (right), for $J=0.3$ and $p=0.5$. 
    The development of a plateau $I_2(A:R)\approx 0$ with increasing system size is associated to a QECC of growing distance.
    (Inset) Contiguous distance $d_\text{cont}$, extracted from $\min \{|A|: I_2(A:R)>1\}$, exhibits a power-law divergence.
    }
    \label{fig:code}
\end{figure}

We start from a fully mixed state on $\tilde{L}$ qubits ($\tilde{L}$ even) and evolve it with dual-unitary Clifford gates and two-qubit Bell measurements as described in the main text. 
A Bell measurement on a pair of qubits $(\tilde{j}, \tilde{j}+1)$ is only allowed within a light cone $|2\tilde{j}+1-\tilde{L}| < 2t$; the dynamics stops after $\tilde{T} \equiv \tilde{L}/2$ steps, so that the final time step involves measurements anywhere in the system (`light cone' boundary conditions). 
We put the final stabilizer state in the `clipped gauge'~\cite{Nahum2017, Li2018}, in which the location of stabilizer endpoints carries a precise physical meaning: given two contiguous regions $A$ and $B$, with $A$ to the left of $B$, the mutual information $I_2(A:B)$ is equal to the number of stabilizers that start in $A$ and end in $B$.
Thus the distribution of stabilizer lengths in the clipped gauge $P(\ell)$ contains important information about the spatial structure of entanglement in the state.
In particular, the mutual information between a contiguous region $A$ and its complement is approximately proportional to the average stabilizer length $\overline{\ell} = \int_0^{|A|} d\ell P(\ell) \ell$. 
At entanglement critical points this distribution develops a power-law tail $P(\ell)\sim \ell^{-2}$, yielding $I_2(A:\bar{A}) \sim \log |A|$. 
This was conjectured to hold into the mixed phase as well (and appears to hold in Haar-random unitary circuits with measurement~\cite{Fan2020}). 
However a tail $P(\ell)\sim \ell^{-\beta}$, $\beta<2$ has been observed in numerical simulations on Clifford circuits, which translates to a \emph{power-law divergent} mutual information between subsystems, $I_2(A:\bar{A})\sim |A|^{2-\beta}$. 
Here we find similar behavior, see Fig.~\ref{fig:code}(a): a power-law tail with exponent seemingly smaller than 2. 
Numerical fits indicate $2-\beta \simeq 0.2$ rather than the value $\simeq 0.38$ conjectured to be universal for Clifford mixed phases; this could be a genuine difference between the models studied, or could reflect a consequence of boundary conditions or of having evolved the system for shorted times (typically one uses $T\gtrsim L$ while the light-cone boundary conditions impose $T = L/2$). 

With the same method one can compute the mutual information between the \emph{reference qubits} $R$ (i.e. an auxiliary system that purifies the initial fully mixed state) and subsystems $A$.
This can be obtained entirely from the final mixed state:
\begin{equation}
    I_2(A:R) = S(A) + S(R) - S(A\cup R)
    = S(A) + S(A\cup \bar{A}) - S(\bar{A}) \;,
\end{equation}
which is entirely independent of $R$, and we have used the fact that $A\cup \bar{A} \cup R$ is in a pure state.
A region $A$ such that $I_2(A:R)\simeq 0$ is one where measurements cannot read-out of the encoded information. In this sense, formation of a QECC (and thus hiding of information in increasingly non-local degrees of freedom) must be accompanied by the emergence of a length scale $d_\text{cont}$ such that $|A|<d_\text{cont} \implies I_2(A:R)\simeq 0$. 
We can define the contiguous code distance  as $d_\text{cont} = \min \{|A|: I_2(A:R)\geq 1\}$. Numerical results in Fig.~\ref{fig:code}(b) show that, for subsystems near the edge, the contiguous distance vanishes -- information about $R$ is immediately gained as qubits are sequentially accessed from the edge. 
This makes sense, as qubits a distance $x$ from the edge participate to the dynamics only for $x$ time steps, and thus do not have time to get strongly entangled with the rest. On the contrary, for regions in the bulk, we see the formation of a plateau in $I_2(A:R)$ vs $|A|$ indicative of a power-law divergent contiguous distance, $d_\text{cont}\sim L^\alpha$.
The exponent appears to be $\alpha\simeq 0.5$.
This exponent, too, was conjectured to take on a universal value $\sim 0.38$ in the mixed phase.
Thus it appears that in these models $\alpha \neq 2-\beta$, and that both may be different from the value $\simeq 0.38$ reported in Ref.~[\onlinecite{Li2020}]. 
A more thorough investigation of this potential discrepancy is an interesting direction for future work.

\bibliography{NoPostselection}